\documentclass[useAMS,usenatbib]{mn2e}
\usepackage{graphicx}
\usepackage{epstopdf}
 \def\thjet{\theta_{\rm j}}
\def\thmax{\theta_{\rm max}} 

\title[joint optical and gravitational wave emissions]{Toward an optimal search strategy of optical and gravitational wave emissions from binary neutron star coalescence.}

\author[]{D.M. Coward$^{1}$\thanks{E-mail:coward@physics.uwa.edu.au}, B. Gendre$^{2}$, P.J. Sutton$^{3}$, E.J. Howell$^{1}$, T. Regimbau$^{4}$,  \newauthor M. Laas-Bourez$^{1}$, A. Klotz$^{5}$, M. Bo\"er$^{6}$, M. Branchesi$^{7,8}$  \\
$^{1}$School of Physics, University of Western Australia, Crawley WA 6009, Australia\\
$^{2}$ASI Science Data Center, via Galileo Galilei, 00044 Frascati (RM), Italy\\
$^{3}$ School of Physics and Astronomy, Cardiff University, The Parade, Cardiff CF24 3AA\\
$^{4}$Dpt. ARTEMIS, Observatoire de la C$\hat{o}$te d'Azur, BP 429 06304 Nice, France\\
$^{5}$ Centre d'\'{e}tude spatiale des rayonnements, Observatoire Midi-Pyr\'{e}n\'{e}es, CNRS,\\ Universit\'{e} de Toulouse, BP 44346, F--31028 Toulouse Cedex 4, France\\
$^{6}$Observatoire de Haute-Provence, F--04870 Saint Michel l'Observatoire, France\\
$^{7}$DiSBeF - Universit\`a degli Studi di Urbino `Carlo Bo', I-61029 Urbino, Italy\\
$^{8}$INFN, Sezione di Firenze, I-50019 Sesto Fiorentino, Italy\\}
\begin{document}
\vspace{-5mm}

\pagerange{\pageref{firstpage}--\pageref{lastpage}} \pubyear{3002}

\maketitle

\label{firstpage}

\begin{abstract}
Observations of an optical source coincident with gravitational wave emission detected from a binary neutron star coalescence will improve the confidence of detection, provide host galaxy localisation, and test models for the progenitors of short gamma ray bursts. We employ optical observations of three short gamma ray bursts, 050724, 050709, 051221, to estimate the detection rate of a coordinated optical and gravitational wave search of neutron star mergers. Model \emph{R}-band optical afterglow light curves of these bursts that include a jet-break are extrapolated for these sources at the sensitivity horizon of an Advanced LIGO/Virgo network. Using optical sensitivity limits of three telescopes, namely TAROT (m=18), Zadko (m=21) and an (8-10) meter class telescope (m=26), we approximate detection rates and cadence times for imaging. We find a median coincident detection rate of 4 yr$^{-1}$ for the three bursts. GRB 050724 like bursts, with wide opening jet angles, offer the most optimistic rate of 13 coincident detections yr$^{-1}$, and would be detectable by Zadko up to five days after the trigger. Late time imaging to $m=26$ could detect off-axis afterglows for GRB 051221 like bursts several months after the trigger. For a broad distribution of beaming angles, the optimal strategy for identifying the optical emissions triggered by gravitational wave detectors is rapid response searches with robotic telescopes followed by deeper imaging at later times if an afterglow is not detected within several days of the trigger.
\end{abstract}

\begin{keywords}
stars -- gamma-ray burst: individual -- gravitational waves -- techniques: miscellaneous -- stars: neutron
\end{keywords}

\section{Introduction}

A multi-messenger approach to gravitational wave detection is one of the most prioritized goals for second generation ground-based gravitational wave (GW) detectors, such as Advanced LIGO
\citep{aligo} and Advanced Virgo \citep[][]{avirgo}.  It allows GW candidates that are too weak to claim detection based on GW data alone to be associated with an optical signal that could provide strong confirmation. \citet{koch93} showed that in principle a joint electromagnetic-gravitational wave (EM-GW) search reduces the GW amplitude detection threshold by a factor of about 1.5. This has the effect of extending the sensitivity horizon distance of GW detectors for a binary neutron star merger (NS-NS), so that the number of potentially detectable GW sources increases by a factor of $\sim 3.4$.\\
The benefits of joint EM-GW searches are significant on many fronts. For example, direct measurement of GWs from a NS-NS late-inspiral and merger provide a means to determine luminosity distance; by combining GW derived luminosity distance measurements with EM redshifts, one can constrain key cosmological parameters. Joint observations offer unprecedented insight into the complex astrophysics that are key to the EM emissions, offering the most complete picture from the strong to weak gravity regime.

One expected EM counterpart of a NS-NS merger is a short gamma ray burst (SGRB), where `short' is defined as $T_{90}<2$s\footnote{The duration in which the cumulative counts increase from 5\% to 95\% above background.}. The popular model for SGRBs is a compact object merger triggering an
explosion causing a burst of collimated $\gamma$-rays \citep{elp+89,npp92, lrg05} powered by accretion onto the newly formed compact
object. SGRBs are believed to be produced by dissipation of kinetic energy of
ultra-relativistic outflow from the central engine with a Lorentz factor of
$\Gamma \sim $100--1000.  The outflow is eventually decelerated by interaction
with interstellar matter to produce a fading x-ray and optical afterglow. After $\Gamma$ decreases to $\Gamma \sim \thjet^{-1}$, where $\thjet$ is the jet opening half angle, the radiation beam is wider than the
outflow, so the afterglow becomes observable from angles greater than $\thjet$.

 Although NS-NS mergers are the favoured progenitor for SGRBs, other scenarios cannot be excluded, i.e. NS-BH mergers \citep{McWilliamsLevin_2011} or magnetar outbursts \citep{Nakar}. The main evidence is based on the association of some SGRBs with an older stellar population, as compared to `long' GRBs, which are associated with massive stellar collapse. Evidence for the origin of SGRBs in the final
merger stage comes both from the
host galaxy types (e.g. Zheng \& Ramirez-Ruiz 2007) and the
measured offsets of GRBs from their host galaxies (e.g. Belczynski
et al. 2006). Kicks imparted to NSs at birth will produce velocities of several hundred km s$^{-1}$, implying that binary inspiraling systems may
occur far from their site of origin. Fong, Berger \& Fox (2009) using Hubble Space Telescope observations to measure SGRB galaxy offsets, find the offset distribution compares favorably with the predicted distribution for NS-NS binaries. However, they do not rule
out at least a partial contribution from other progenitors systems.

\subsection{GW triggered SGRB afterglow search}\label{rates}
A number of laser interferometric GW detectors have now reached their
design sensitivities and have been operating as a global array, coordinating
with electromagnetic observations through triggered follow-ups.
These include the LIGO\footnote{http://www.ligo.caltech.edu/}
detectors based at Hanford and Livingston in USA, the
Virgo\footnote{http://www.virgo.infn.it/} detector in
Italy and the GEO\,600\footnote{www.geo600.uni-hannover.de}
detector in Germany. The LIGO and Virgo detectors are undergoing
a series of upgrades towards Advanced configurations that will produce an order of magnitude improvement in
sensitivity. Advanced
LIGO\footnote{www.ligo.caltech.edu/advLIGO/} and Advanced
Virgo\footnote{www.cascina.virgo.infn.it/advirgo/} are expected to be
operational by 2015.

Because the coalescing binary NSs are expected to radiate GWs in the sensitivity band of Advanced LIGO/Virgo, coincident GW-EM observations of SGRBs will determine if the engine is a NS-NS or NS-BH binary merger. Furthermore, the rates of EM and coincident GW detections could constrain the distribution of jet collimation angles of SGRBs, crucial for understanding energetics \citep{ber07}. This is possible because the binary inclination angle to the line of sight is a GW observable\footnote{In practice, for sources not associated with host galaxies, the inclination angle has a strong degeneracy with distance, particularly for angles less than 45$^{\rm{o}}$.}. A direct consequence of collimation is that the rate of (both long and short) GRB afterglows should be higher than those observed as prompt bursts. SGRBs afterglows observed `off-axis' without a prompt counterpart have been termed `orphan afterglows'. There has not been a definitive discovery of an orphan afterglow, despite both dedicated searches to $m=23$ and using archived data \citep{tot02,rau06}. The non-detection could partly be attributed to the small flux of off-axis afterglows, compared to on-axis ones that can be observed as early as seconds after the prompt burst and in some cases while the high-energy prompt emission is still occurring \citep{Klotz09}. 

The prospects might seem bleak for detecting off-axis afterglows. This is the case for an all-sky search, but a GW observation of a NS-NS inspiral collapses the search area from all-sky to the error ellipse (angular sky resolution) of the GW detectors. For such events, the angular resolution of a detector network depends on GW source strength, orientation of the binary axis, and on the geometrical configuration of the network \citep{wen_2010}. Error ellipses of the order of tens to a few square degrees can be achieved through triangulation of time differences in signal arrival times at various detectors in a network \citep{gursel_89,Fairhurst_09}.

The sensitivity distance, $D_{s}$, for sources uniformly distributed in orientation and sky location is approximated by $D_{H}/2.26$, were $D_{H}$ is the horizon distance in Mpc at which an optimally-orientated, overhead source can be detected with a signal-to-noise ratio of 8 \citep{lsc1}. We note that although $D_{\mathrm{H}}$ is based on the sensitivity of single detector, non-optimal effects such as non-Gaussian, non-stationary detector noise allow the approximation to be assumed for an Advanced LIGO/Virgo network \citep{lsc1}. Significantly, the value $D_{s}$ of Advanced LIGO/Virgo for a NS-NS coalescence is smaller than the average distance to the observed SGRB population. This implies that EM emissions from NS-NS coalescences triggered by Advanced LIGO/Virgo will be brighter than the {\it Swift}\footnote{http://swift.gsfc.nasa.gov/docs/swift/swiftsc.html} triggered GRB emissions. \citet{nnt10} showed that a galaxy ranking procedure could identify the host galaxy 75-95\% of the time out to 100 Mpc for 5 images taken by narrow field and wide field telescopes respectively. Their method depends on galaxy survey completeness, so may not be applicable for the Advanced GW detector network searches. We discuss the issues that will affect the localisation of SGRBs in Section {\ref{disc}.

To estimate the optical flux of a SGRB as the EM counterpart of a NS-NS coalescence, we use the plausible estimates of \citep{lsc1} for the Advanced LIGO/Virgo sensitivity distances and detection rates of NS-NS coalescences. Taking a rate density of NS-NS coalescences $\sim 10^{-6}$\,Mpc$^{-3}$yr$^{-1}$ and $D_{\mathrm{H}}= 445\,\mathrm{Mpc}$, they find $D_s\approx 200$ Mpc and a detection rate $R_{\mathrm{det}}\sim 40$ yr$^{-1}$ . This rate could potentially be increased by considering the improved signal to noise ratio for a coincident GW and optical search. The estimated increase in signal to noise ratio is about 1.5, assuming a narrow coincidence window, but the optical afterglows may not be imaged until hours after the GW trigger. Nonetheless, the afterglows should be relatively bright at the distances we are considering so it is possible that light curves could be extracted and extrapolated to earlier times. Hence, we assume the sensitivity distance increases by a factor of 1.5, to 300 Mpc, so that $R_{\mathrm{det}}\sim$ 135 yr$^{-1}$.


In this Letter we investigate the temporally varying optical brightness of SGRBs at the Advanced LIGO/Virgo sensitivity distance to NS-NS coalescences, from the first hour of the EM emission to about a hundred days later. We use optical data for three SGRBs, GRB 050724, GRB 050709 and GRB 051221, that show evidence for collimated emission and are localised. Assuming that all NS-NS mergers produce jets, we use these data combined with reasonable estimates of the NS-NS coalescence rate detected by a GW detector network to determine the fraction of events that could be potentially detected as both on and off axis bursts. For definiteness, we use the sensitivity limits of two robotic telescopes, TAROT \citep[$m=18$, see][]{Klotz08}, and Zadko \citep[$m=21$, see][]{cow10}, that have participated in the optical follow-up of LIGO/Virgo GW triggers as part of the LOOC UP program during the 6th Science Run \citep{LOOCUP}. We also include a much deeper sensitivity limit of $m=26$, representative of an $(8-10)$m class telescope such as the VLT\footnote{The Very Large Telescope, see http://www.eso.org/public/teles-instr/vlt.html}. Finally we discuss the main issues that need addressing to interpret and optimize the science return from joint optical and GW searches.


%
\vspace{-5mm}
\section{Optical afterglow model and coincident rates}
\subsection{Observations}\label{obs}
In order to constrain the detection rate, we require localisation (including redshift), beaming angles and the optical flux values for our sample of bursts. We use three SGRBs that have estimates for these parameters: namely GRB 050709, GRB 050724, and GRB 051221A.\newline

\noindent{\bf GRB 050709}: From comparison of X-ray and optical data, a jet break is present in the optical at about 10 days after the burst (Fox et al. 2005). On the other hand, Watson et al. (2006) claimed that the light curves were not displaying such a break. We note however that they excluded one optical data point within their fit, arguing it was coincident with a late X-ray flare. However, the data point they excluded was 9.8 days after the burst, compared to 16 days for the X-ray flare. We assume the explanation of Fox et al. (2005), noting that the detection of the jet-break is supported by only one data point.\newline
{\bf GRB 050724}: From radio and near infrared data, Panaitescu (2006) and Berger et al. (2005), claimed evidence of a jet break about 1 day after the burst. However, the main feature of the jet-break is its achromacity (Rhoads 1999), and the X-ray data do not feature such a break (Gruppe et al. 2006). The X-ray light curve is consistent with no jet break up to 22 days after the event, or $\thjet>25$\degr.\newline
{\bf GRB 051221A}: The detection of the jet break was observed in X-ray only (Soderberg et al. 2006). A jet-break is clearly visible in the light curve at about 5 days post-burst.\\

We note that determination of the jet opening angle from the break-time is strongly sensitive on the model parameters used. In the previous cases, the model was always the same (a forward shock fireball expanding with kinetic energy \emph{E} in a medium of constant density \emph{n}), where $\theta \propto (n/E)^{1/8}$. However, $n$ and $E$ are uncertain in all bursts (see e.g. Panaitescu 2006 for the values of $n$ in the cases of GRB 050709 and GRB 050724). Hence, opening angle values reported in these papers are taken as indicative values, and we use them as a guide (this is why we do not perform a k-correction for cosmological effects, as the correction is small compared to the uncertainties of the beaming angles).
\vspace{-5mm}
 \subsection{Afterglow model and rates}

 For GRB 050724, GRB 050709 and GRB 051221, we define $F_{11 \mathrm{hr}}$ using $R$-band fluxes at 11 hr from \citet{Nys09}, but scaled to a source distance of 300 Mpc (the mean sensitivity distance of the GW search).
Equation (1) scales the flux using a power law index $\alpha_1$ until jet-break time $t\le t_{j,0}$, when the beamed emission starts to expand sideways, and by the luminosity distance to the source $d_{\mathrm{L}}(z)$:
\begin{equation}
F(t,z) = \left(\frac{t}{11 \mathrm{hr}}\right)^{\alpha_1} F_{11 \mathrm{hr}} \left(\frac{d_{\mathrm{L}}(z)}{ D_s}\right)^2\,.
\label{eq1}
\end{equation}
The above equation can be used to estimate the $R$-band flux at times before the jet break. To model the SGRB light curve at post jet-break times, we employ a smoothly joined broken power law (Beuermann et al. 1999),
\begin{equation}
F(t) = F_j \left[\left(\frac{t}{t_{j}}\right)^{-\alpha_1 n} +   \left(\frac{t}{t_{j}}\right)^{-\alpha_2 n} \right] ^{-1/n},
\label{F2}
\end{equation}
\noindent where $F_j$ is the flux at the jet break time $t_{j}$, $\alpha_1$ and $\alpha_2$ are the pre-break and post-break light-curve slopes, and \emph{n} scales the sharpness of the break. The beaming angles and break times for GRB 050724, GRB 050709, GRB 051221 are (25\degr, 22d), (14\degr, 10d), (7\degr, 5d) respectively with corresponding power law indices of $\alpha_1$ and $\alpha_2$  $(-1.5,-2)$, $(-1.25,-2.83)$ and $(-1,-2)$.

Table 1 shows the derived parameters with the extrapolated $R$-band magnitude at 1 hr post burst at 300 Mpc. We point out that the optical data used in the references to derive the beaming angle and break times is uncertain, and we do not account for optical bumps and flares that can be significant, especially at early times. Nonetheless, it is clear from Table 1 that if one of the well localised SGRBs occurred within $D_s$, and was on-axis, it would be bright at early times and easily detected by modest aperture telescopes.

To calculate the rate of triggered detection of on-axis afterglows requires accounting for the beaming angle $\thjet$. Equation (\ref{eq3}) calculates the number of afterglows seen per yr for a certain $\thjet$ assuming a LIGO/Virgo detection rate of $R_{\mathrm{det}}\sim135$yr$^{-1}$ (see section \ref{rates})
\begin{equation}
R_{\mathrm {on}} = R_{\mathrm{det}} [1-
\mathrm{cos} (\thjet)]\,.
\label{eq3}
\end{equation}

The rate of off-axis bursts is determined by the maximum angle, $ \thmax$, away from the jet center that an off-axis observer could see the afterglow. This angle depends critically on the flux limit of the telescope $ F_{\rm lim}$, $F_j$ and $\thjet$. For off-axis detection, the main constraint is the off-axis emission only becomes visible after the jet-break time, which may be days after the prompt burst. \cite{tot02}, assuming the uniform jet model, show that $ \thmax$ can be expressed as

\begin{equation}
 \thmax= \left[ 2^{-(3+\delta)} \frac{ F_j }{ F_{\rm lim} }
          \right]^{-1/(2\alpha_2)} \, \thjet \; ,
\label{eq:theta-max}
\end{equation}
where $\alpha_2$ is the post-break optical decay index and $\delta$ is a numerical factor $\sim 1$. $\thjet$ can be replaced with $\thmax$ in equation (\ref{eq:theta-max}) to estimate the SGRB off-axis detection rate;
\begin{equation}
R_{\mathrm {off}}(\thmax) = R_{\mathrm{det}} [1-
\mathrm{cos} ( \thmax)] - R_{\mathrm {on}} .
\end{equation}\label{eq4}
We note that if $\thmax \le \thjet $ only on-axis afterglows will be visible at any time, and the rate is determined by $\thjet$.

\begin{table}
 \caption{The main observed and derived parameters of GRB 050724, GRB 050709 and GRB 051221. See Section \ref{obs} for caveats and uncertainties in $\thjet$. }
 \label{symbols}
 \begin{tabular}{@{}lccccccc}
  \hline
  GRB & log E$_{\mathrm{iso}}$
        & $z$ & R-mag (1 hr)$^\dagger$ & $t_{j}$ (d) & $\thjet(\degr)$  \\
        050724  & 50.21 & 0.26 & 12.7 & $<22$ & 25  \\
        050709 & 49.06 & 0.16 & 17.2 & 10 & 14  \\
        051221 & 50.95 & 0.55 & 13.7 & 5 & 7  \\
  \hline
  \hline
\end{tabular}

 \medskip
 $^\dagger$ magnitudes are converted from flux (Jy) to the AB magnitude system using m$_{\mathrm{AB}}$ = -2.5 log$(F) + 8.9$ at a source distance of 300 Mpc \newline
\end{table}
\vspace{-5mm}
\section{Results}
Using the above relations and light curve characteristics for GRB 050724, GRB 050709 and GRB 051221, we extrapolate the light curves beyond the jet-break times to constrain detection limits, rates and cadence times using the sensitivities of TAROT, Zadko and an $(8-10)$m class telescope. Figure 1 shows the temporal evolution of the three $R-band$ light curves using equations (1) and (2) at a source distance of 300 Mpc, and published values for the decay indices. The three curves are quite different, GRB 050724 and GRB 051221, are relatively bright at early times, and can be seen from days to some tens of days by meter class telescopes.

Table 2 shows  $\thmax$, the detection rates $R_{\mathrm {on}}$ and $R_{\mathrm {off}}$, and the maximum possible times that an on-axis burst would be visible for the three telescopes. We find that GRB 050724 like bursts are detectable at a relatively high rate. This is because the initial beaming angle is large $>25\degr$, so more of the afterglows can be seen on-axis. Conversely, the brightest afterglow at the post break time, GRB 051221, is the least likely to be detected because of the small $\thjet \sim 7\degr$. It is apparent that both TAROT and Zadko are unlikely to detect off-axis afterglows from these SGRBs, but a telescope capable of deep imaging to $m=26$ could detect an additional 5 afterglows yr$^{-1}$ for GRB 051221 like events.

%

Table 2 also shows the maximum time, $t_{\mathrm {max}}$, that the telescopes could detect the SGRB afterglows. This sets the limit on the cadence times for imaging. GRB 051221 has the brightest afterglow and is potentially detectable the longest time; $t_{\mathrm {max}}\sim11$d for Zadko. Given that the GW error ellipse is of order degrees in size, identification of a transient is more feasible for this afterglow type. Unfortunately, they occur at a rate of 1 yr$^{-1}$, and given optical selection effects (see section \ref{disc} for a discussion) may be missed altogether. GRB 050724 like events, occurring at an optimistic rate of 13 yr$^{-1}$, would be detectable up to 5d by Zadko. This would allow time for surveying degree size fields and multiple telescopes at different longitudes to perform follow-up imaging.


Our results in Table 2, based on a broad distribution of beaming angles, suggest that the optimal strategy for identifying the optical emissions triggered by gravitational wave detectors is through initial rapid response searches with robotic telescopes, followed by deeper imaging at later times if an afterglow is not detected within several days of the trigger.

\begin{figure}
\includegraphics[scale=0.4]{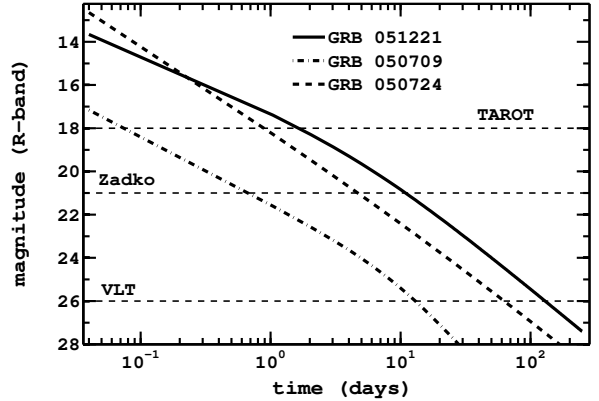}
\caption{Three model light curves using equation (\ref{F2}) for GRB 050724, GRB 050709 and GRB 051221 extrapolated to a source distance of 300 Mpc, the horizon limit for the Advanced LIGO/Virgo detector network. The beaming angles and break times for the model bursts are (25\degr, 22d), (14\degr, 10d), (7\degr, 5d) respectively. Power law indices before and after the breaks are $(-1.5,-2)$, $(-1.25,-2.83)$ and $(-1,-2)$ respectively. The horizontal dashed lines from bottom to top are the approximate sensitivities for an (8-10)m class telescope, Zadko Telescope (1m) and TAROT (0.25m) respectively. The maximum on-axis times for visibility are shown in Table \ref{tab2}.
 } \label{fig1}
\end{figure}

\begin{table}
 \caption{Equations (4) is used to calculate the maximum off-axis angle that the three SGRBs could be observed assuming limiting magnitudes of TAROT, Zadko and an $(8-10)$m class telescope such as VLT. Equations (3) and (5)
 are used to calculate both on and off-axis rates of SGRB afterglows associated with GW emission form NS-NS coalescences detected by a 3-detector network of GW detectors. The maximum time, $t_{\mathrm {max}}$ for detecting the optical afterglows for the three telescope sensitivities is also shown. The median and most optimistic on-axis detection rates are 4 and 13 yr$^{-1}$ respectively. }
 \label{tab2}
 \begin{tabular}{@{}lcccccccc}
  \hline
 GRB & Telescope &   $\thmax$
         & $R_{\mathrm {on}}$  & $R_{\mathrm {off}}$  &  $t_{\mathrm {max}}$  \\
         & & (deg) & (yr$^{-1}$) & (yr$^{-1}$) & (d) \\
       050724 & TAROT & $\thmax<\thjet$ & 13 & 0 & 1 \\
       & Zadko & - & 13 & 0 & 5  \\
        & VLT & $\thmax\sim\thjet$ & 13 & 0 & 60  \\
        \hline
         050709  & TAROT & $\thmax<\thjet$ & 4 & 0 & 0.08 \\
         & Zadko &  - & 4 & 0 & 1  \\
         & VLT & - & 4 & 0 & 13 \\
         \hline
         051221  & TAROT & $\thmax<\thjet$ & 1 & 0 & 2 \\
            & Zadko & - & 1 & 0 & 11  \\
         & VLT & 18  & 1 & 5 & 130   \\
  \hline

  \hline
 \end{tabular}
 \vspace{1mm}
The maximum off-axis viewing angle, $\thmax$, is not shown for afterglows with $\thmax<\thjet$. This also implies a non-detection of an off-axis afterglow. The rates are upper limits and do not account for Galactic extinction and crowded fields.
\end{table}
\vspace{-5mm}
\section{Discussion}\label{disc}
The first attempts for a triggered search of the optical counterparts of NS-NS coalescences using GW detectors are just commencing. There are many uncertainties and issues that will need careful consideration for these types of searches. Firstly, our results show that the coincident detection rate depends critically on the beaming angle distribution. For nearly isotropic optical emission, similar to GRB 050724, the coincident rates are very promising and will improve the confidence of the GW detection and provide much needed localisation. Even non-detections of optical emissions for high signal-to-noise ratio NS-NS GW candidates is interesting. Non-detections of a statistically significant sample would constrain the SGRB beaming angle distribution, or show that SGRBs are not linked to NS-NS coalescences. Both implications are critical to our understanding of the progenitors of SGRBs.

To fully use joint optical and GW observations requires understanding and accounting for selection effects that have historically plagued SGRB optical observations. Typical selection effects include Galactic extinction and crowded star-fields. There is also the problem of host galaxy extinction, although this is expected to be a less of a problem given that at least some of the afterglows are observed off-set from the hosts. From the current optical follow-up attempts of {\it Swift} triggered bursts, it is clear that a significant fraction of SGRBs that are apparently on-axis have not been observed in the optical at all. This can partly be attributed to the much greater distances of the {\it Swift} bursts, compared to a GW triggered search. We have based our calculations on a sensitivity distance of $\sim 200$ Mpc (non-coincidence rate), but note that GW wave sources with smaller angles of inclination will be detected at greater distances. Preliminary studies suggest that this bias can increase the median distance to $\sim 269$\,Mpc for binaries with inclination angles less than 25\degr. A numerical study is planned to investigate how this bias will effect the rate of coincident detections.

Another important issue for the joint searches is the large errors in the GW source localisation, which can extend to some tens of degrees. This is a significant problem given the first observed GRB optical afterglows required small error boxes of arc-minute size. To counter this, the GW triggered search strategy will use the estimated horizon distance of the detector network to reduce the number of potential host galaxies, as opposed to a `blind' error box that extends to cosmological distances. This technique was demonstrated in the recent LIGO/Virgo S6; the telescope pointing strategy used a galaxy weighting taking into account the mass and the distance of catalogued galaxies. A detailed description and analysis of the S6 coincidence searches will be published as a LIGO Scientific Collaboration paper.

 Another problem that manifests with large coincidence error boxes is the increasing chance of detecting false coincident optical transients. False coincident sources may include supernovae, flare stars, variable active galactic nuclei and even Earth orbiting space debris. Fortunately, some of these sources can be excluded in the analysis because of the sensitivity distance of GW searches and the expectation that the strongest GW sources will be associated with catalogued host galaxies. Another possibility for optimizing the coincidence search is to data mine archived images from very wide field optical surveys, such as {\it SkyMapper} \footnote{http://www.mso.anu.edu.au/skymapper/}, and into the future the planned {\it Large Synoptic Survey Telescope} \footnote{http://www.lsst.org/lsst}.

Given the above practical difficulties, an observed association of an optical transient with a NS-NS coalescence triggered by Advanced LIGO/Virgo is challenging. But, the science pay-off for such a discovery is enormous and provides motivation to address the issues discussed here in more detail. Now is the time to determine the optimal strategies for optical follow-up in readiness for the more sensitive GW searches in the following years. To accomplish this will require a more comprehensive understanding of optical selection effects, the false alarm rate expected from SGRBs within the error ellipses of GW networks, and techniques to improve the localisation of the host galaxy.

\vspace{-5mm}
\section*{Acknowledgments}
D.M. Coward is supported by an Australian Research Council Future Fellowship. P. J. Sutton was supported in part by STFC grant 500704.

\vspace{-5mm}
%
%

\label{lastpage}


\vspace{-5mm}

\begin{thebibliography}{99}

\bibitem[Abadie et al.(2010)]{lsc1} Abadie J., et al., 2010, Class. Quant. Grav., 27, 173001

\bibitem[\protect\citeauthoryear{Acernese et al.}{2009}]{avirgo} Acernese F. et al., 2009, Advanced Virgo baseline design, \textit{Virgo Internal Note} VIR-0027A-09
\bibitem[Beuermann et al (1999)]{be99}Beuermann K., et al. 1999, A\&A, 352, L26
\bibitem[Belczynski et al (2006)]{bel06}Belczynski K., Perna R., Bulik, T., Kalogera V., Ivanova N., \& Lamb D. Q., 2006, ApJ, 648, 1110
\bibitem[Berger et al (2005)]{ber05}Berger, E., et al. 2005, Nature, 438, 988
\bibitem[Berger et al (2007)]{ber07}Berger E. et al. 2007, ApJ, 664, 1000
\bibitem[Bloom et al. (2006)]{blm06}Bloom J. S., et al. 2006, ApJ, 638, 354
\bibitem[Coward et al. (2010)]{cow10}Coward D.M. et al., 2010, PASA, 27, 331

\bibitem[Eichler et al.(1989)]{elp+89} Eichler D. et al.\ 1989, Nature, 340, 126
\bibitem[Evans et al.(2007)]{ebp+07} Evans, P.~A., et al.\ 2007, A\&A, 469, 379

\bibitem[\protect\citeauthoryear{Fairhurst}{Fairhurst}{2009}]{Fairhurst_09}
Fairhurst S.,  2009, New Journal of Physics, 11, 123006

\bibitem[Fong, Berger \& Fox(2010)]{fong10}Fong W., Berger E., \& Fox D.B., 2010, ApJ, 708, 9

\bibitem[Fox et al., (2005)]{fox05}  Fox et al., 3005, Nature, 437, 845

\bibitem[Gehrels et al., (2005)]{geh05}Gehrels, N., et al. 2005, Nature, 437, 851
\bibitem[\protect\citeauthoryear{Harry et al.}{2010}]{aligo} Harry G.~M. et al., 2010, Class. Quant. Grav., 27, 084006

\bibitem[Gruppe et al., (2006)]{grup06}Gruppe et al., 2006, ApJ 653, 462

\bibitem[\protect\citeauthoryear{G\"ursel \& Tinto}{G\"ursel \&
  Tinto}{1989}]{gursel_89}
G\"ursel Y.,  Tinto M.,  1989, Phys. Rev. D, 40, 3884


\bibitem[\protect\citeauthoryear{Kanner at al.}{2008}]{LOOCUP}
Kanner J. et al.,  2008, Class. Quant. Grav., 25, 184034

\bibitem[Klotz et al.(2008)]{Klotz08}
Klotz, A. et al. 2008, PASP, 120, 1298

\bibitem[Kochanek \& Piran (1993)]{koch93} Kochanek C.S., Piran, T. 1993, ApJ, 417, L17

\bibitem[Klotz et al.(2009)]{Klotz09}
Klotz, A. et al. 2009, ApJ, 697, L18

\bibitem[Lee et al.(2005)]{lrg05} Lee W.~H., Ramirez-Ruiz, E., Granot, J.\ 2005, ApJL, 630, L165
\bibitem[Malesani et al. (2007)]{Mal07}Malesani et al. 2007, A\&A,  473, 77

\bibitem[\protect\citeauthoryear{{McWilliams} \& {Levin}}{{McWilliams} \&
  {Levin}}{2011}]{McWilliamsLevin_2011}
{McWilliams} S.~T.,  {Levin} J.,  2011, submitted to Nature (astro-ph.HE/1101.1969)

\bibitem[Nakar(2007)]{Nakar} Nakar E.,  2007, Physics Reports, 442, 166
\bibitem[Narayan et al.(1992)]{npp92} Narayan R., Paczynski B., Piran, T.\ 1992,ApJL, 395, L83

\bibitem[Nuttall \& Sutton(2010)]{nnt10}Nuttall L., Sutton P., 2010, Phys. Rev. D., 82, 102002

\bibitem[Nysewander, Fruchter \& Pe'er (2009)]{Nys09}Nysewander M., Fruchter A. S., Pe'er A., 2009, ApJ, 701, 824

\bibitem[Panaitescu(2006)]{pan06}Panaitescu, A. 2006, MNRAS, 367, L42

\bibitem[Rau, Greiner \& Schwarz(2006)]{rau06}Rau A., Greiner J., Schwarz R., 2006, A\&A 449, 79

\bibitem[Rhoads(1999)]{rho99}Rhoads, J. E. 1999, ApJ, 525, 737

\bibitem[Soderberg et al. (2006)]{Sod06}Soderberg A.M., et al., 2006, ApJ, 650, 261

\bibitem[Totani \& Panaitescu(2002)]{tot02}Totani T., Panaitescu A., 2002, ApJ, 576, 120

\bibitem[Watson et al.(2006)]{wat06}Watson et al. 2006, A\&A, 454, L123

\bibitem[\protect\citeauthoryear{Wen \& Chen}{Wen \& Chen}{2010}]{wen_2010}
Wen L.,  Chen Y.,  2010, Phys. Rev. D, 81, 082001

\bibitem[Zheng \& Ramirez-Ruiz (2007)]{z07}Zheng Z. \& Ramirez-Ruiz E., 2007, ApJ, 665, 1220


%
%

\end{thebibliography}
\end{document}